\documentclass[twocolumn]{aastex631}

\newcommand\nickel{$^{56}$Ni }
\newcommand\edot{$\dot{E}_\mathrm{expl}$}

\newcommand\sawadap{\citep{2019ApJ...886...47S}}
\newcommand\sawadat{\citet{2019ApJ...886...47S}}

\newcommand\suwat{\citet{2019MNRAS.483.3607S}}
\newcommand\prevt{\citet{2023MNRAS.518.1818I}}

\usepackage{threeparttable}

\submitjournal{ApJ}

\usepackage{amsmath}

\graphicspath{{./}{figures/}}

\shorttitle{Updating the $^{56}$Ni Problem in CCSNe}
\shortauthors{Sawada \& Suwa}

\begin{document}


\title{Updating the $^{56}$Ni Problem in Core-collapse Supernova Explosion}

\correspondingauthor{Ryo Sawada}
\email{ryo@g.ecc.u-tokyo.ac.jp}

\author[0000-0003-4876-5996]{Ryo Sawada}
\affiliation{Department of Earth Science and Astronomy, Graduate School of Arts and Sciences, The University of Tokyo, Tokyo 153-8902, Japan}

\author[0000-0002-7443-2215]{Yudai Suwa}
\affiliation{Department of Earth Science and Astronomy, Graduate School of Arts and Sciences, The University of Tokyo, Tokyo 153-8902, Japan}
\affiliation{Center for Gravitational Physics and Quantum Information, Yukawa Institute for Theoretical Physics, Kyoto University, Kyoto 606-8502, Japan}

\begin{abstract}
Details of the core-collapse supernova (CCSN) explosion mechanism still need to be fully understood. 
There is an increasing number of successful examples of reproducing explosions in multidimensional hydrodynamic simulations, but subsequent studies pointed out that the growth rates of the explosion energy $\dot{E}_\mathrm{expl}$ of these simulations are insufficient to produce enough $^{56}$Ni to match observations. 
This issue is known as the `$^{56}$Ni problem' in CCSNe.
Recently, however, some studies have suggested that this $^{56}$Ni problem is derived from the simplicity of the explosion model.
In response, we investigate the effect of the explosion energy growth rate $\dot{E}_\mathrm{expl}$ on the behavior of nucleosynthesis in CCSNe in a more realistic model. 
We employ the 1D Lagrangian hydrodynamic code, in which we take neutrino heating and cooling terms into account with the light-bulb approximation. 
We reiterate that, consistent with previous rebuttal studies, there is the $^{56}$Ni problem: Although $^{56}$Ni is synthesized to almost the same mass coordinate independent of $\dot{E}_\mathrm{expl}$, some of the innermost material in the low-$\dot{E}_\mathrm{expl}$ model failed to escape, leading to a shift in the innermost mass coordinate of the ejecta to the outer positions.
Comparing our results with observations, we find that while modern slow explosions can, in principle, reproduce observations of standard Type II SNe, this is not possible with stripped-envelope SNe. 
Our finding places a strong constraint on the explosion mechanism. 
There are significant differences in the progenitor structures and the explosion mechanism between Type II and stripped-envelope SNe.
\end{abstract}

\section{Introduction}\label{sec:intro}
Radioisotope \nickel is an important product in supernova nucleosynthesis, which drives supernova (SN) brightness. \nickel decays into $^{56}$Co, and then into $^{56}$Fe. This nuclear decay chain powers the light curve of SNe,
and thus, \nickel masses of SNe have been estimated with reasonable accuracy from the light curve \citep[see, e.g., ][]{1982ApJ...253..785A,2003ApJ...582..905H}.\footnote{For Type-II SNe, the tail luminosity provides \nickel mass, assuming the complete trapping of $\gamma$-rays produced from the nuclear decay. For Type-I SNe, on the contrary, the peak luminosity has often been used, assuming that it should be equal to the instantaneous energy deposition rate by the nuclear decay, so-called Arnett rule \citep{1982ApJ...253..785A}. Core-collapse SNe in the latter category is called stripped-envelope SNe \citep[SE-SNe;][]{2009ARA&A..47...63S}.}
On the other hand, the amount of synthesized \nickel is sensitive to the temperature $T$, the density $\rho$, and the number of electrons per nucleon (electron fraction) $Y_e$, i.e., explosion property and pre-SNe core structure \citep[e.g.,][]{1995ApJS..101..181W,1996ApJ...460..408T,2002RvMP...74.1015W}.
These two factors, that is, the amount of \nickel synthesis can be accurately estimated from observations and strongly reflect the explosion's innermost nature, suggest the following. \nickel is the best probe to constrain an aspect of the SN explosion mechanism accurately \citep[e.g.,][]{2009MNRAS.394.1317M,2015MNRAS.451..282S}.

Details of the explosion mechanism of core-collapse supernovae (CCSNe) are not yet fully understood. 
The most promising scenario is the delayed neutrino-driven explosion \citep{1985ApJ...295...14B}. 
While this scenario had once not been reproduced by numerical simulations, the situation has brought substantial progress over a few decades.
Now, there is an increasing number of successful examples of reproducing explosions in multidimensional hydrodynamic simulations, with a detailed neutrino transport \citep[see, e.g., ][and references therein]{2015ApJ...807L..31L,2016MNRAS.461L.112T,2017MNRAS.472..491M,2018ApJ...865...81O,2019ApJ...873...45G,2021ApJ...915...28B,2021Natur.589...29B,2022arXiv221112675B}. 
Although the details now depend on the numerical methods and physical approximations employed in each simulation, there seems to be a general understanding that the explosion succeeds by the growth of the hydrodynamic instability over a sufficient time.
Indeed, most, if not all, of those state-of-the-art simulations, have shown a slow increase of explosion energy, and the growing rate of the explosion energy is typically $\dot{E}_\mathrm{expl}=\mathcal{O}(0.1)$ Bethe s$^{-1}$ (1 Bethe$\equiv 1 \times 10^{51}$ erg), especially for 3D simulations.

However, recent several studies have shown that to reproduce the typical observed mass of \nickel by the explosive nucleosynthesis in the ejecta, the growth rate of the explosion energy of $\dot{E}_\mathrm{expl}=\mathcal{O}(1)$ Bethe s$^{-1}$ is required in several methods \citep{2019ApJ...886...47S,2019MNRAS.483.3607S,2022ApJ...931..153S}.
\citet{2019ApJ...886...47S} found the inverse-correlation between \nickel yield and explosion energy growth rate \edot~ by 1D simulations with the simple thermal-bomb modeling and post-processing detailed-nucleosynthesis, and \citet{2019MNRAS.483.3607S} also came to the same conclusion by conducting hydrodynamic simulations with an approximate neutrino heating model that self-consistently follows core-collapse and shock-revival. 
\citet{2022ApJ...931..153S} also confirmed this trend, using the same method as \citet{2019ApJ...886...47S}, but modeled for individual objects to reduce observational uncertainties.
If these results are correct, the current multi-D simulations, which give explosion energy growth rates of $\dot{E}_\mathrm{expl}=\mathcal{O}(0.1)$ Bethe s$^{-1}$, would be observationally unfavorable.
We refer to this issue as the nickel mass problem (`\nickel problem,' hereafter) in this paper.
However, this \nickel problem is still under some debate.

In particular, \prevt~just recently pointed out the most obvious question to the  \nickel problem.
\prevt~used the same method as \citet{2019ApJ...886...47S}, with simple thermal injection modeling and post-processing detailed nucleosynthesis, but scrutinized the treatment of initial conditions.
In the recent slow explosion scenario, the pre-SN star experiences sufficient gravitational contraction just before the successful explosion.
They found that the correlation between \nickel yield and explosion energy growth rate \edot~ is the result of ignoring this initial collapse.
They argued that this correlation disappears when the initial collapse is included and also that further initial collapse inversely results in more \nickel being synthesized in slower explosions.

Their arguments also apply to \citet{2019ApJ...886...47S} and \citet{2022ApJ...931..153S}, but not to \citet{2019MNRAS.483.3607S}.
\citet{2019MNRAS.483.3607S} solved self-consistently the core collapse and shock revival with the light-bulb scheme and found this correlation even though they took into account the initial collapse phase. 
This result is inconsistent with the conclusion of \prevt.
Note that \citet{2019MNRAS.483.3607S} performed no detailed nucleosynthesis calculations. Instead, they estimated the \nickel amount simply by the temperature of hydrodynamic simulations.
Therefore, we perform hydrodynamic and detailed nucleosynthesis calculations in this study. 
This study aims to clarify the detailed picture of how \nickel synthesis occurs in the current CCSN explosion scenario.
By clarifying this picture, we also expect to explain the origin of the differences between the two studies and, by extension, the cause of the \nickel problem itself.

In this paper, therefore, we simulate one-dimensional hydrodynamics in the light-bulb scheme as in \suwat, then perform detailed nucleosynthesis in a post-process manner.
The goal of this study is to present a detailed picture of \nickel nucleosynthesis in CCSNe with self-consistent explosion modeling.
Furthermore, we aim to sort out the controversial \nickel problem.
In Section \ref{sec:method}, we describe our simulation methods, the progenitor models, and post-processing analysis. Our results are summarized in Section \ref{sec:result}.
In Section \ref{sec:discus}, we revisit the \nickel problem through a detailed comparison of our results and observations, and discuss the uncertainties involved. 
We conclude in Section \ref{sec:summary}.

\section{Simulation Methods}\label{sec:method}

Following the computational setup performed in \citet{2019MNRAS.483.3607S}, 
we employ a 1D Lagrangian Newtonian hydrodynamic code based on {\tt blcode}.\footnote{This code is a prototype code of {\tt SNEC} \citep{2015ApJ...814...63M}, and available from https://stellarcollapse.org}
Basic equations under a spherically symmetric configuration, as we perform in this paper, are given as follows: 
\begin{align}
\cfrac{\partial r}{\partial M_{r}}&=\cfrac{1}{4\pi r^2\rho} \label{eq:basic1}~,\\
\cfrac{D v}{D t}&=-\cfrac{GM_{r}}{r^2}-4\pi r^2\cfrac{\partial P}{\partial M_{r}} \label{eq:basic2}~,\\
\cfrac{D \epsilon}{D t}&=-P\cfrac{D}{D t}\left(\cfrac{1}{\rho}\right)+\mathcal{H}-\mathcal{C} \label{eq:basic3}~,
\end{align}
where $r$ is the radius, $M_{r}$ is the mass coordinate, $t$ is time, $\rho$ is the density, $v$ is the radial velocity, $P$ is pressure, $\epsilon$ is the specific internal energy, and $D/Dt \equiv \partial/\partial t + v_r\partial/\partial r$ is the Lagrangian time derivative. 
The artificial viscosity of \cite{1950JAP....21..232V} is employed to capture a shock. 
The system of equations (\ref{eq:basic1})-(\ref{eq:basic3}) is closed with the Helmholtz equation of state \citep{2000ApJS..126..501T}, which describes the stellar plasma as a mixture of arbitrarily degenerate and relativistic electrons and positrons, black-body radiation, and ideal Boltzmann gases of a defined set of fully ionized nuclei, taking into account corrections for the Coulomb effects. 

In this work, neutrino heating and cooling are added by a light-bulb scheme.
In the light-bulb scheme, neutrino cooling is given as a function of temperature, and neutrino heating is a function of the radius with parameterized neutrino luminosity. 
The heating term $\mathcal{H}$ and the cooling term $\mathcal{C}$, terms in Equation \eqref{eq:basic3} are assumed to be
\begin{align}
\mathcal{H}&=1.544 \times 10^{20} ~\mathrm{erg}~\mathrm{g}^{-1} ~\mathrm{s}^{-1} \nonumber\\
&\times \left(\cfrac{L_{\nu_e}}{10^{52}\mathrm{MeV}}\right)
 \left(\cfrac{r_{\nu_e}}{100\mathrm{km}}\right)^{-2}
 \left(\cfrac{T_{\nu_e}}{4.0\mathrm{MeV}}\right)^2
\label{eq:neut1}~,\\
\mathcal{C}&=1.399 \times 10^{20} \,\ \mathrm{erg}\,\ \mathrm{g}^{-1} \,\ \mathrm{s}^{-1} 
\times \left(\cfrac{T}{2.0\mathrm{MeV}}\right)^6
\label{eq:neut2}~.
\end{align}
Here, we fix the neutrino temperature as $T_{\nu_e}=4~\mathrm{MeV}$. We take into account these terms only in the post-shock regime. 
We modified the inner boundary conditions so that the innermost mass shell does not shrink within $50$ km from the center to mimic the existence of a proto-neutron star (PNS).
Also, the light-bulb scheme in this study tends to overestimate the neutrino-driven wind from the PNS surface at the post-explosion phase because it keeps giving a constant neutrino luminosity. 
Therefore, in this study, we consider the mass coordinate that experienced $r<200$ km as the neutrino-driven wind and separate it from the ejecta.

The numerical computational domain contains $1.5 M_\odot$ and uses a 1500 grid with a mass resolution of $10^{-3}M_\odot$. We set the inter boundary at $M_{s/k_b=4}-0.5 M_\odot$ for each pre-explosion star.
Each mass coordinate captures the time evolution of the hydrodynamic quantities, so that nucleosynthesis calculations are performed as a post-processing analysis with this trajectory. 
We calculate a reaction network of 640 nuclear species with the {\tt torch}  code \citep{1999ApJS..124..241T}.

\begin{figure}
\centering
\includegraphics[width=0.49\textwidth]{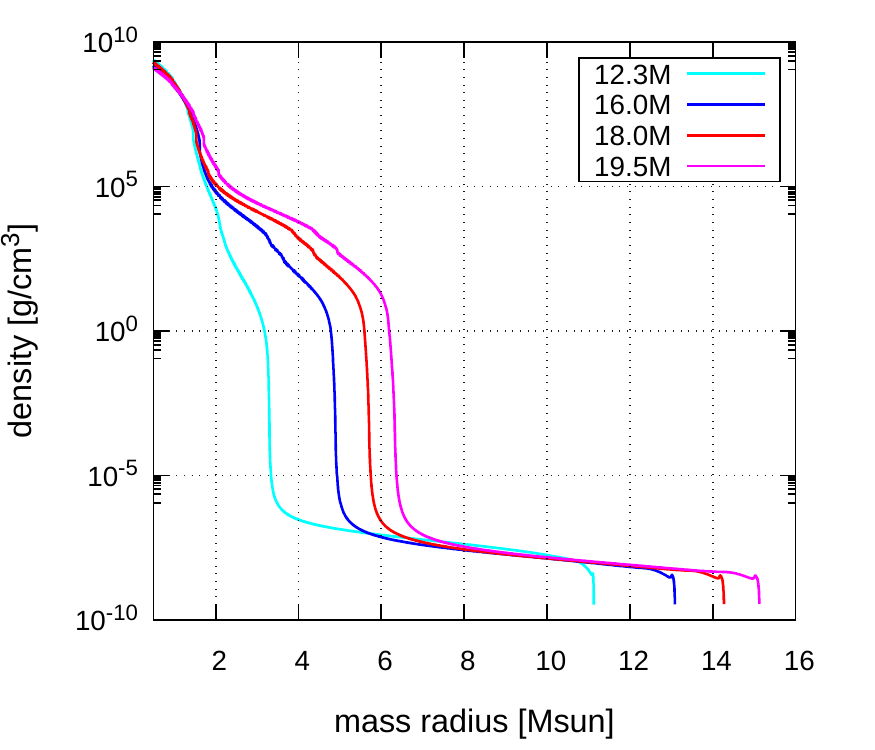}
\includegraphics[width=0.49\textwidth]{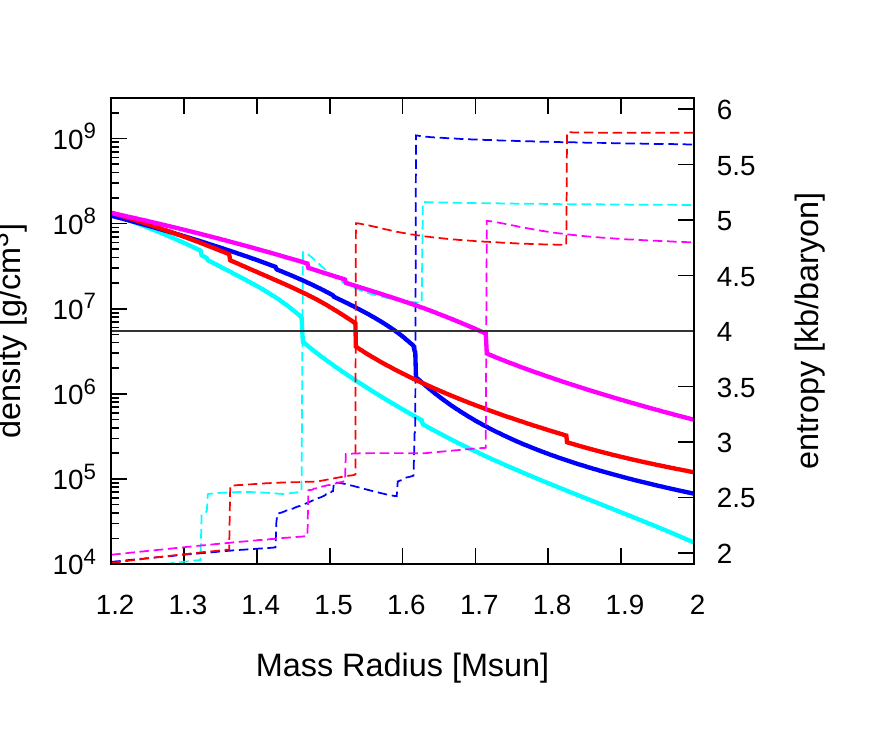}
\caption{Density structure as a function of the enclosed mass for the considered progenitors with $M_\mathrm{ZAMS}=12.3M_\odot$ (cyan line), $16.0M_\odot$ (blue line), $19.7M_\odot$ (red line), and  $21.0M_\odot$ (magenta line), and its details with the entropy per nucleon.} 
\label{fig:progenitor}
\end{figure}

The initial conditions adopted in this study are a subset of 
non-rotating stars with solar metallicity, which evolved from the main sequence to the onset of iron-core collapse, as published by \citet{2018ApJ...860...93S}. 
The physics of this set of progenitors was discussed in detail in this literature.
Figure \ref{fig:progenitor} shows the density structures of the progenitor as a function of the enclosed mass, and its details with the entropy per nucleon.

\section{result}\label{sec:result}

\subsection{Overview of the explosion dynamics}\label{sec:result-expl}
\begin{figure}
\centering
  \includegraphics[width=0.49\textwidth]{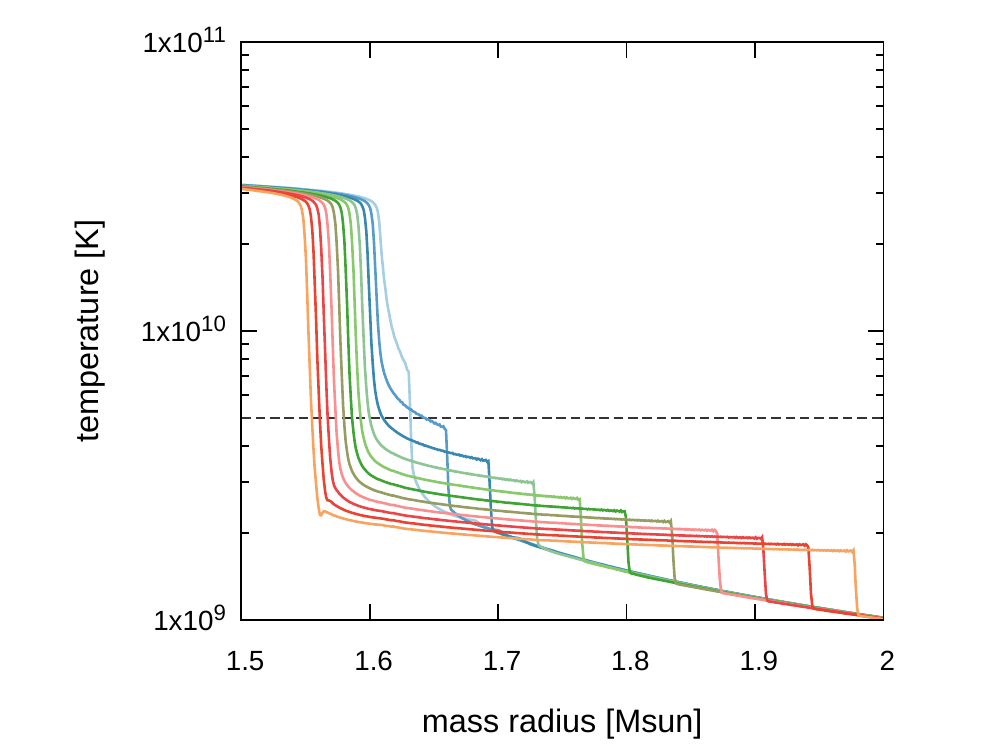}
  \includegraphics[width=0.49\textwidth]{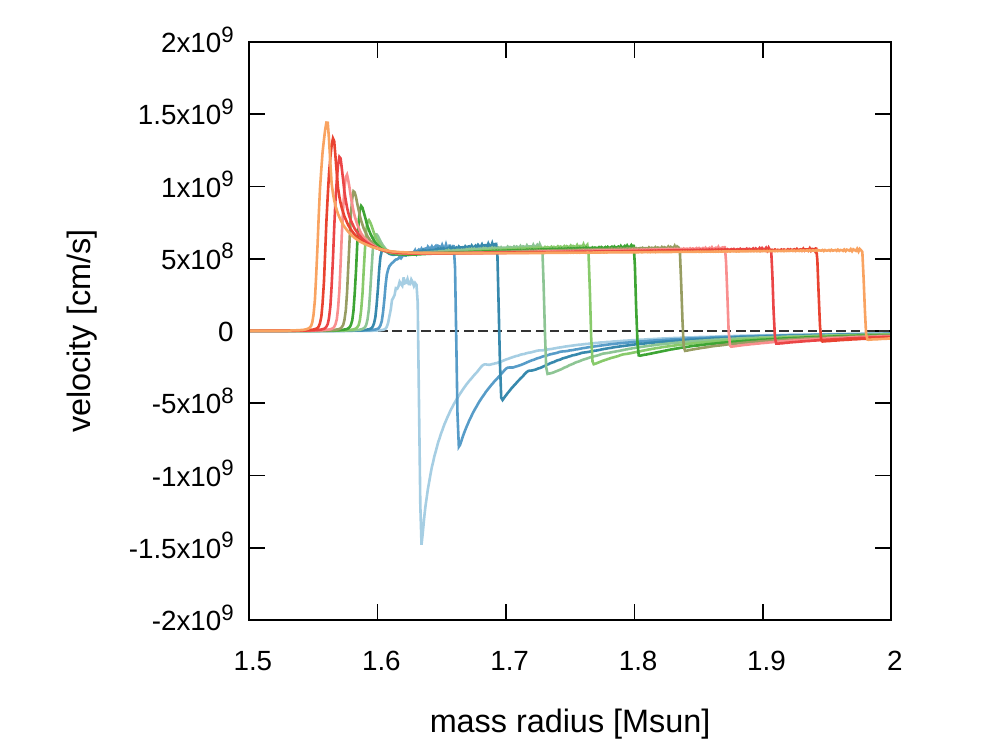}
\caption{Time evolution of the velocity (top) and the temperature (bottom) as a function of the mass coordinate for model 16.0$M_\odot$. In both panels, each snapshot time corresponds to approximately every $0.1$ seconds from $0.5$ seconds to $1.5$ seconds from the start of the simulation. The gray line corresponds to $T=5\times 10^9$ K.} 
\label{fig:hyd}
\end{figure}

\begin{figure}
\centering
  \includegraphics[width=0.49\textwidth]{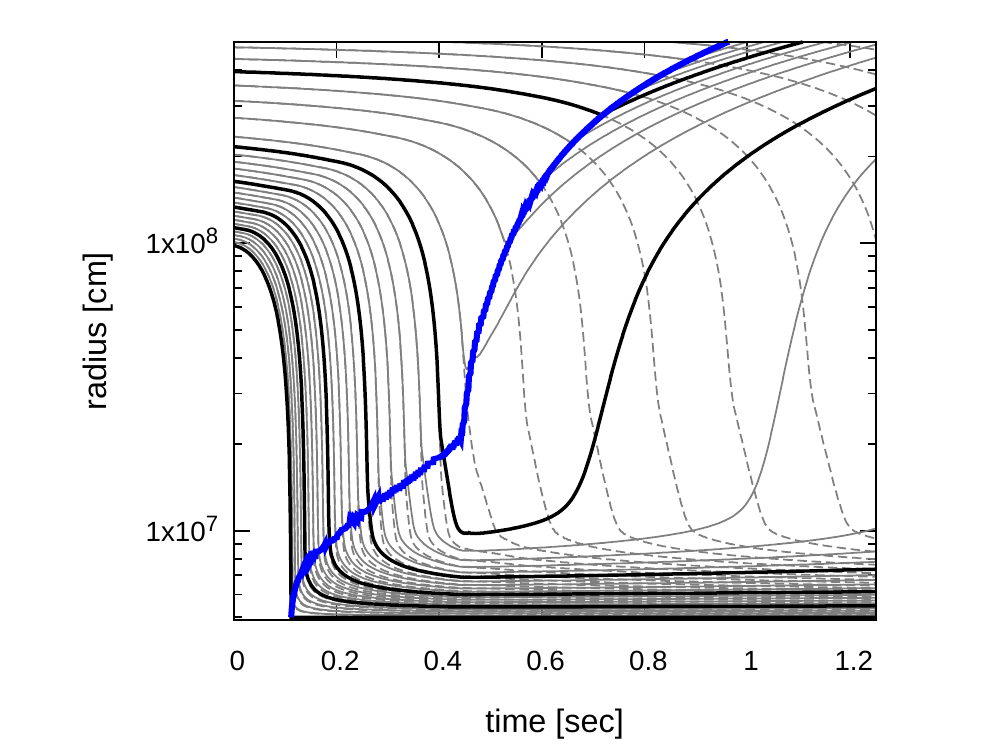}
\caption{Radius evolution of Lagrangian mass shells with time for the explosion ($L_\nu = 3\times10^{52}$ erg s$^{-1}$) and non-exploding model ($L_\nu = 0$ erg s$^{-1}$) of 16.0$M_\odot$. The thick black solid lines are the mass shells, spaced in steps of 0.1 $M_\odot$, and the thin gray solid/dashed lines are spaced in steps of 0.02 $M_\odot$. The difference between the dotted and solid lines corresponds to the explosion and non-exploding models, respectively. The blue line marks the shock radius of the explosion model.} 
\label{fig:l3e52}
\end{figure}

Figure \ref{fig:hyd} shows the time evolution of radial velocity and temperature as a function of the mass coordinate for the model $16.0M_\odot$. 
We first use an example of a model with $M_\mathrm{ZAMS} = 16.0 M_\odot$ throughout this section. 
From the velocity figure, it can be seen that the shock begins to propagate outward from the point where the silicon/oxygen (Si–O) layer ($\approx1.62 M_\odot$) accretes onto the shock wave, due to the rapid decrease in ram pressure \citep{2009ApJ...694..664M,2016ApJ...816...43S}. 
From the temperature figure, we can confirm that the post-shock temperature of the ejecta is spatially almost constant so we define the shock temperature as the temperature of the material just behind the shock wave.

Figure \ref{fig:l3e52} shows that the mass shell until the arrival of the shock in the explosion model is consistent with its behavior in the non-exploding model. 
In other words, we can confirm that the behavior of the mass shell up to the arrival of the shock is independent of the explosion detail.
Thus, by overlaying the shock evolution on the trajectory of the mass shell in the non-exploding model, we can compare several models at once to see where the shock impacts each of the mass shells.

In the following subsections, we present the results focusing on the effect of the explosion energy growth rate {\edot} on \nickel nucleosynthesis.
The results are summarized in Table \ref{tbl:result}. 
These yields consist only of unbound \nickel by gravity as determined by a 10-second simulation.
We first use an example of a model with $M_\mathrm{ZAMS} = 16.0 M_\odot$ throughout Sections \ref{sec:result-hyd} and \ref{sec:result-nucl}. 

\subsection{Hydrodynamics and \nickel Synthesis Region}\label{sec:result-hyd}

\begin{figure}
\centering
  \includegraphics[width=0.49\textwidth]{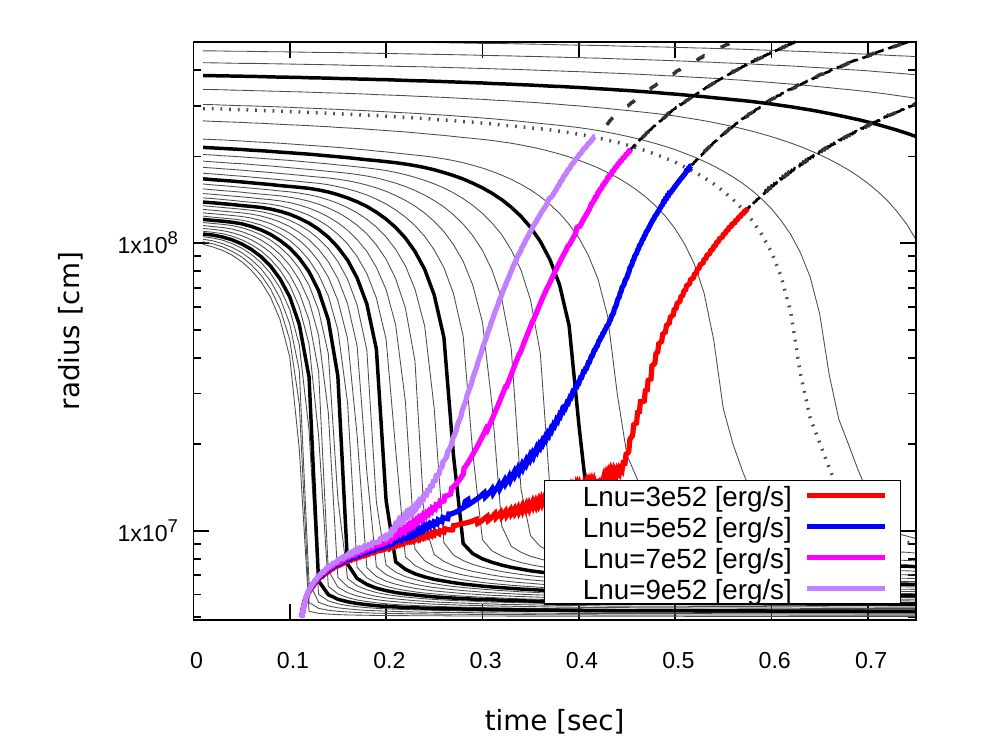}
\caption{The time evolution of the shock radius in models $L_\nu = 3,~5,~ \mathrm{and}~ 7\times10^{52}$ erg s$^{-1}$ with the mass shell trajectory in the unexploded model, on the time-radius plane.} 
\label{fig:comp-shock}
\end{figure}

Figure \ref{fig:comp-shock} shows the time evolution of the shock for the models $L_\nu = 3,~5,~ \mathrm{and}~ 7\times10^{52}$ erg s$^{-1}$ and the trajectory of the mass shell in the unexploded model.
In each model, the time evolutions of the shock are shown by colored lines for the range where the shock satisfies $T_9>5$ ($T_9\equiv T/10^9$ K), and by black dashed lines for the range where $T_9 < 5$.
We refer to the mass coordinate that can spread at the shock temperature of $T_9 \approx 5$ as $M_{T_9=5}$. 
We find that in all models $M_{T_9=5}$ is near the mass coordinate with the enclosed mass $\approx1.65 M_\odot$, which is indicated by the dotted line in Figure \ref{fig:comp-shock}.
More detailed values are given in Table \ref{tbl:result}, and this trend is almost universal, independent of the progenitor models.
In all models, we find that $M_{T_9=5}$ is near the mass coordinate with an enclosed mass of approximately $1.65M_\odot$, as indicated by the dotted line in Figure \ref{fig:comp-shock}. 
More detailed values are given in Table \ref{tbl:result}, and this trend is nearly universal and independent of the progenitor models. 

Qualitatively, this can be understood by using a zero-order approximation to estimate the shock radius at which the shock temperature is $T_9=5$.
When applying a simple fireball model in which the region behind the shock wave is uniform and dominated by radiation pressure \citep[e.g.,][]{2002RvMP...74.1015W}, we can estimate the following relation between the temperature $T$, the shock radius $r_\mathrm{sh}$ and the explosion energy $E_\mathrm{expl}$ as follows:
\begin{equation}
  E_\mathrm{expl}
  =\cfrac{4\pi}{3}r_\mathrm{sh}^3(t)~aT^4~,
\end{equation} 
where $a$ is the radiation constant. 
Then with $E_\mathrm{expl}(t)\equiv10^{51}$ ergs, the radius with $T_9=5$ ($r_{_{T_9=5}}$) can be estimated as follows: 
\begin{equation}
    r_\mathrm{_{T_9=5}} \approx 3.6\times 10^8 (E_\mathrm{expl}/10^{51})^{1/3}~\mathrm{cm}~.
\end{equation} 
This estimated radius is classically well-known \citep[e.g.,][]{2002RvMP...74.1015W,2013ARA&A..51..457N}.
If we consider the time evolution of $E_\mathrm{expl}(t)=\dot{E}_\mathrm{expl}\cdot t$, this classical radius is satisfied with an adequate large \edot. 
However, at the shock velocity $V_\mathrm{sh}=10^9$ cm s$^{-1}$, it takes less than 1 second to reach this radius.
In other words, if the case of $\dot{E}_\mathrm{expl}\lesssim 1$ Bethe s$^{-1}$, it takes a few seconds to reach 1 Bethe, and obviously, the radius of $T_9=5$ will be small.
In fact, from Figure \ref{fig:comp-shock}, we can confirm that even in this simulation, the radius of $T_9=5$ is reduced in the case of low-\edot.
But at the same time, the time evolution of the shock radius is also slower down for lower-\edot~models, and the mass shell falls more inward due to collapse. Eventually, the `mass coordinate' of $M_{T_9=5}$ seems to be approximately the same regardless of the \edot.

This result suggests a very interesting trend.
\nickel is synthesized mainly by complete Si burning at $T \gtrsim 5\times10^9$ K (see in detail in Appendix \ref{sec:nickel} and \citealt{1973ApJS...26..231W}). 
Thus, this hydrodynamical result suggests that the outermost mass coordinates, where \nickel is primarily synthesized, are insensitive to the explosion energy growth rate \edot.
To confirm this trend in more detail, we next discuss the results of nucleosynthesis calculations.

\begin{figure}
\centering
  \includegraphics[width=0.49\textwidth]{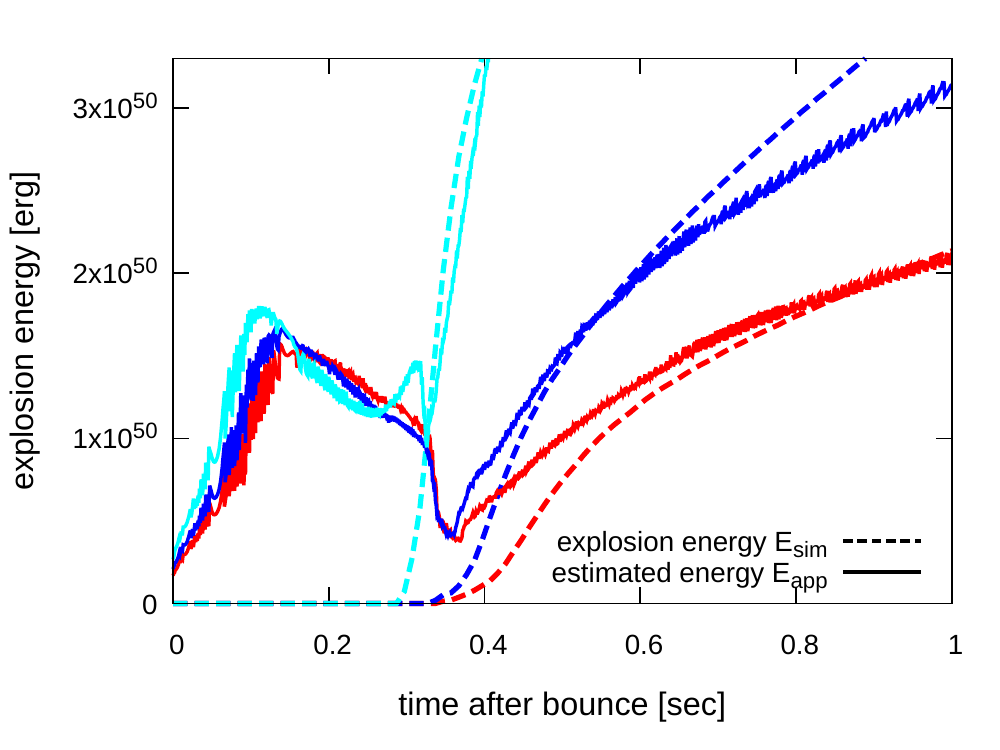}
\caption{Comparison of the explosion energy in the simulation $E_\mathrm{sim}$ (dashed line) with the estimated energy in the fireball approximation $E_\mathrm{app}$ (solid line), which comes from Eq \eqref{eq:fireball} in models $L_\nu = 2,~3,~ \mathrm{and}~ 5\times10^{52}$ erg s$^{-1}$. The horizontal axis is the post-bounce time.}
\label{fig:expl-ene}
\end{figure}

Although this is not relevant to the main focus of this paper,
we show in Figure \ref{fig:expl-ene} for reference the comparison of the explosion energy in the simulation $E_\mathrm{sim}$ with the estimated energy in the fireball approximation $E_\mathrm{app}$.
The explosion energy $E_\mathrm{sim}$ in the hydrodynamical simulation is defined as the integral of the sum of specific internal, kinetic, and gravitational energies over all zones, in which it is positive.
The estimated energy in the fireball approximation $E_\mathrm{app}$ is given by the following equation using only the shock radius $r_\mathrm{sh}$ and shock temperature $T$:
\begin{equation}
  E_\mathrm{app}
  =\cfrac{4\pi}{3}r_\mathrm{sh}^3(t)~aT^4f(T_9)~, \label{eq:fireball}
\end{equation} 
where $f(T_9)=1+(7/4)\cdot T_9^2/(T_9^2+5.3)$ is a correction term to account for both radiation pressure and non-degenerate electron-positron pairs \citep[e.g.,][]{1999ApJ...516..381F}.
As Figure \ref{fig:expl-ene} shows, this simple estimation is able to reproduce the explosion energy of the simulation with good enough accuracy. This supports the validity of the above discussion, and also suggests that thermal energy is dominant in the early phases of the explosion.

\subsection{Nucleosynthesis: Distribution of \nickel synthesis}\label{sec:result-nucl}

\begin{figure*}[t]
\gridline{\fig{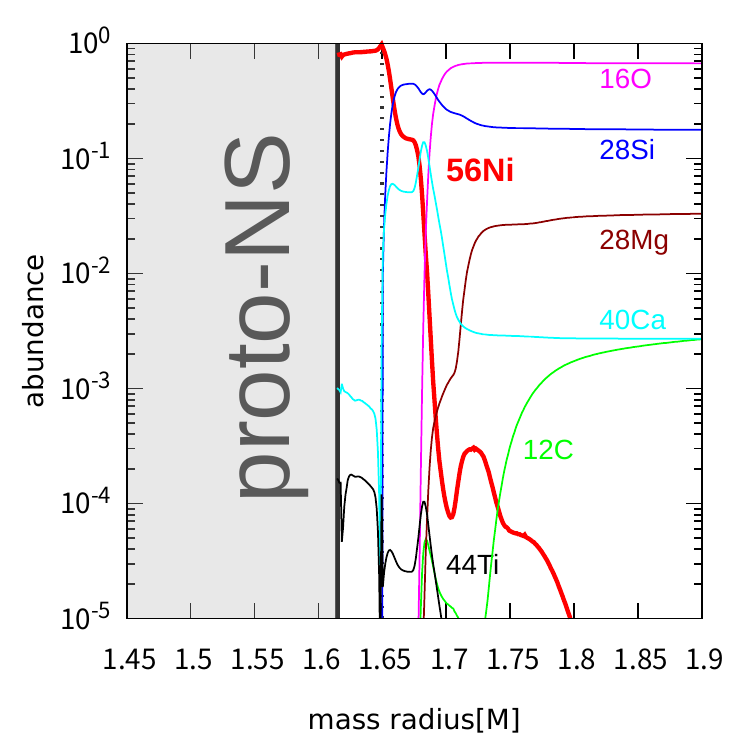}{0.3333\textwidth}{(a)}
          \fig{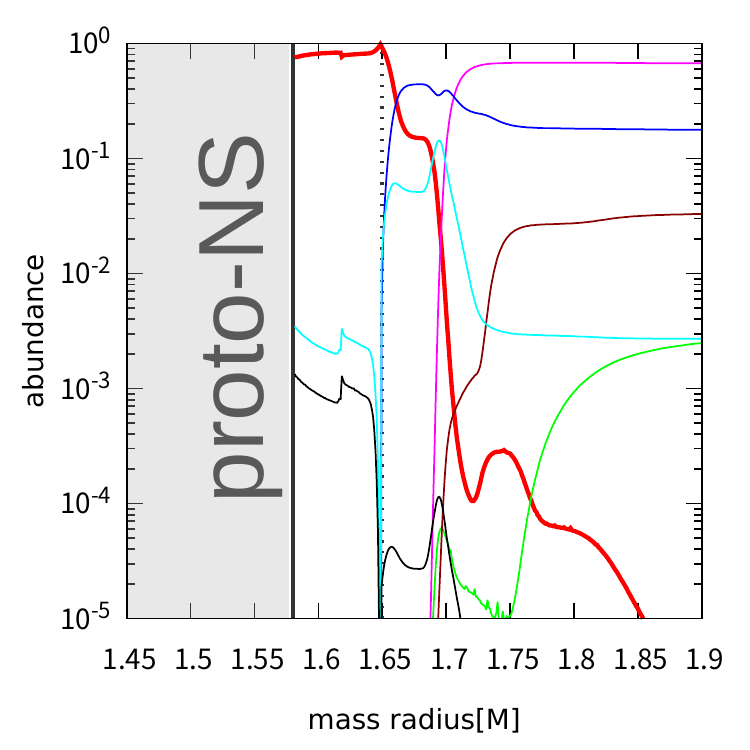}{0.3333\textwidth}{(b)}
          \fig{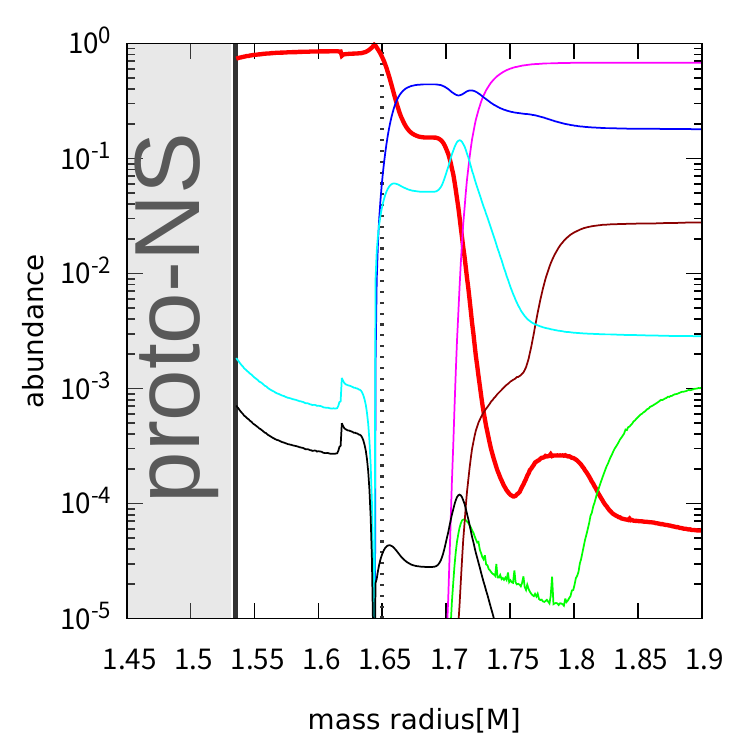}{0.3333\textwidth}{(c)}
          }
\caption{Abundance distribution as a function of the enclosed mass $Mr$, 
for (a) $L_\nu = 3\times10^{52}$ erg s$^{-1}$, (b) $L_\nu = 5\times10^{52}$ erg s$^{-1}$, and (c) $L_\nu = 7\times10^{52}$ erg s$^{-1}$. 
All the models here are with $16.0 M\odot$ of \citet{2018ApJ...860...93S}.
In all panels, the vertical dotted grey line indicates the location of the mass shell with an enclosed mass $1.65M_\odot$.} 
\label{fig:abundance}
\end{figure*}

Figure \ref{fig:abundance} shows the abundance distribution as a function of the mass coordinate, for $L_\nu = 3,~5,~ \mathrm{and}~ 7\times10^{52}$ erg s$^{-1}$ with $16.0 M_\odot$.
As shown in Figure \ref{fig:abundance}, focusing specifically on $^{56}$Ni, we can confirm that the outermost mass radius, where \nickel is primarily synthesized, is in a similar position independent of the explosion energy growth rate \edot. 
In Table \ref{tbl:result}, we show the outermost mass radius where \nickel is largely synthesized (here, we define it as $X(^{56}\mathrm{Ni})>0.5$).
However, at the same time, the innermost mass radius, which is gravitationally unbounded, depends strongly on the explosion energy growth rate \edot.

\section{Discussion: Update `Ni Problem'}\label{sec:discus}
\begin{figure}
\centering
  \includegraphics[width=0.49\textwidth]{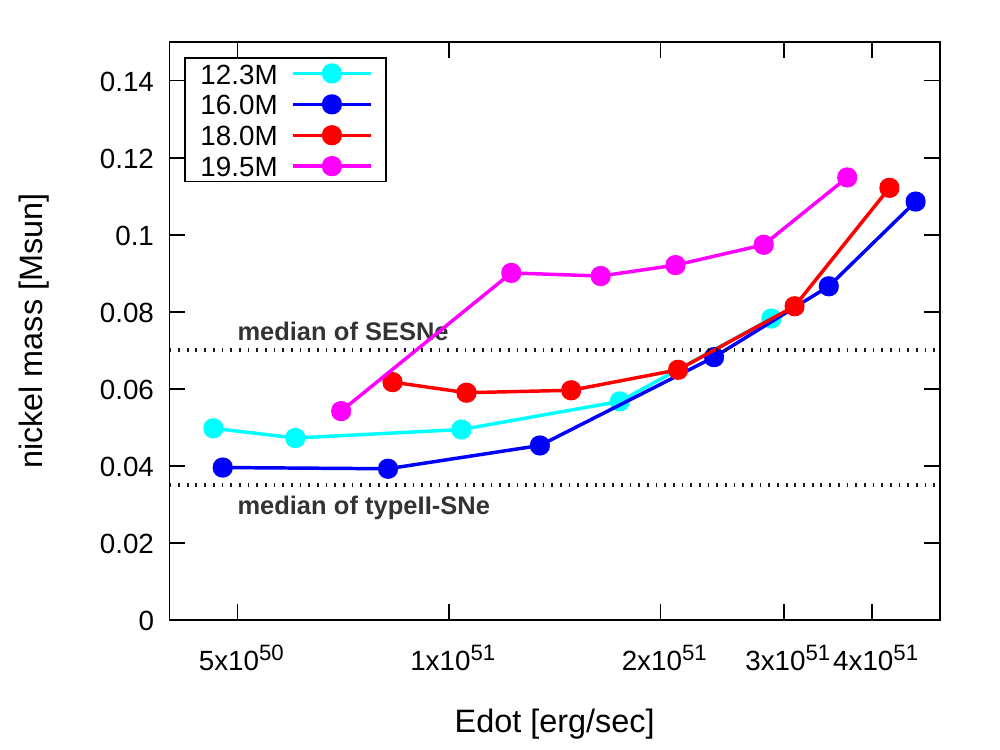}
\caption{The amount of \nickel as a function of the growth rate of the explosion energy, \edot. The gray line indicates a typical value of \nickel, $0.07M_\odot$.} 
\label{fig:edot-nickel}
\end{figure}

Figure \ref{fig:edot-nickel} shows the synthesized amount of \nickel as a function of the explosion energy growth rate \edot.
It can be clearly seen that there is a decreasing trend of the synthesized amount of \nickel toward decreasing \edot.
The reason for this trend is explained in section \ref{sec:result-nucl}, but this figure tells us that the same trend is generally observed regardless of the mass and structure of the progenitor.

For comparison with observations, in Figure \ref{fig:edot-nickel}, we adopted two typical values based on a recent systematic survey for more than 300 events of CCSNe; $0.07M_\odot$\footnote{The $\sim0.07M_\odot$ is often adopted as a typical value obtained for well-studied nearby SNe is on average \citep[e.g., SN 1987A, SN 1994I, SN 2002ap; ][]{1989ARA&A..27..629A,1994ApJ...437L.115I,2002ApJ...572L..61M} and we adopt this value in previous studies. However, in this study, we clearly mention here that we do not use $0.07M_\odot$ in the context of the typical for nearby SNe because we consider observational constraints from recently updated large-scale observational data.}
as the median estimated from stripe-envelope supernovae (SE-SNe) and $0.03M_\odot$ from Type-II SNe \citep[][]{2021MNRAS.505.1742R,2022arXiv220905552R}.
Note that the figure plots the synthesized amount of \nickel; not all \nickel can be ejected.
In other words, the figure shows the maximum amount of \nickel that can be ejected by each CCSN model, and if the calculated mass of \nickel is larger than the observed value, then the model can reproduce the observed value.
First, compared to the median value of Type II supernovae $0.03M_\odot$, even a modern slow explosion ($\dot{E}_\mathrm{expl}\lesssim1$ Bethe s$^{-1}$ ) provides enough amount of \nickel to reproduce the observations.
On the other hand, compared to the SE-SNe median of $0.07M_\odot$, a very rapid explosion of \edot$\gtrsim2$ Bethe s$^{-1}$ is required to reproduce this value.
This translates to a time scale of $t\lesssim0.5$ seconds to the typical explosion energy $\sim1.0$ Bethe, and this timescale is very difficult to reproduce with current multi-D self-consistent calculations.

Here we discuss a few caveats in this problem as follows.

\begin{enumerate}
   \item
{\bf[Observation of Type-II SNe]}\\
The typical \nickel mass of canonical-CCSNe has been extensively discussed by large-scale observations in recent years.
In particular, Type II SNe, when volume-limited, account for nearly $\sim60\%$ of the observed CCSNe \citep[e.g.,][]{2011MNRAS.412.1441L,2021ApJ...908..143J}.
Recently, Type II SNe have been found to have lower median nickel masses than SE-SNe \citep[e.g.,][]{2019A&A...628A...7A}, confirming that this is not due to observational bias \citep{2021ApJ...922..141O}.
Furthermore, the observed kinetic energy is also found to have a lower median value than the classical typical value \citep[$\sim0.6$ Bethe; ][]{2022A&A...660A..41M}.
These facts also support the possibility that the `slow' explosion results in the current state-of-the-art simulations are relatively consistent with standard Type II SNe.
However, nickel synthesis and explosion energy ($M_\mathrm{Ni}\approx0.03M_\odot$ and $E_\mathrm{expl}\sim0.6$ Bethe) still remain important benchmarks for multidimensional self-consistent simulations, and it should be checked whether they are truly achieved.
And, another important point is that this is only a statement of the median.
According to \citet{2021ApJ...922..141O}, the fitting function for the cumulative histogram for observed \nickel masses of each CCSNe is $f(x)= \tanh{(14.60\times x)}$ \footnote{This function is obtained by fitting non-linear least squares of observed reports of Type II SNe with a sample size of 115 events \citep{2021ApJ...922..141O}.} as a variable of observed \nickel masses. This roughly implies that more than 20\% of the Type II supernovae synthesize \nickel above $0.075M_\odot$.
While $0.03M_\odot$ is a somewhat explainable value, this value is challenging to reproduce in multi-D self-consistent simulations.
So, we need to explain and reproduce the high \nickel objects that will exist to some extent.

   \item
{\bf[Multidimensional effect]}\\
How the amount of synthesized \nickel changes in a multi-D explosion model is one of the issues to be discussed.
Since this model is a 1D model and explodes only with thermal energy as shown in Figure \ref{fig:expl-ene}, 
the temperature should be higher than the multi-D model, especially considering the geometric structure and the change to kinetic energy in the non-radial direction \citep{2019MNRAS.483.3607S}.
Therefore, we should note that the same \edot~in a multi-D model would have less \nickel than in a 1D model. 
In fact, with the exception of particular model results \citep[][ discussed next]{2021ApJ...915...28B}, the multi-D self-consistent simulation has even more difficulty with \nickel synthesis than the estimate of this study \citep[e.g.,][]{2022arXiv221112675B}.
Therefore, for the same synthesis conditions, the 1D model gives a robust maximum limit on the volume and the amount of \nickel synthesis. 
The additional \nickel amount newly occurring due to the multi-D effect will be discussed next.

   \item
{\bf[Additional mechanism to add \nickel]}\\
Another possibility for an additional \nickel, and one of the most often cited candidates for a solution to this problem, is the `outflow' from the PNS surface for several seconds of the post-explosion phase \citep[e.g.,][]{2017ApJ...842...13W,2021ApJ...921...19W}.
Recent detailed simulations have predicted proton-rich ejecta in the post-explosion `outflow' \citep[e.g.,][]{2016ApJ...818..123B}.
In particular, \citet{2021ApJ...915...28B} have observed the downflow/outflow system that results in a smooth and efficient transition from the incoming flow to the outgoing flow, with the outflow providing \nickel to $\lesssim 0.05 M_\odot$.
However, we already found that the contribution of such replenishment is small for regular CCSNe explosions \citep{2021ApJ...908....6S}.
That is, this outflow system is part of an `energetic' model of the state-of-the-art simulations that succeeds in producing sufficient amounts of \nickel, and it is debatable whether this outflow system contributes to canonical-CCSNe explosions.

\begin{figure*}[t]
 \begin{center}
  \includegraphics[width=0.98\textwidth]{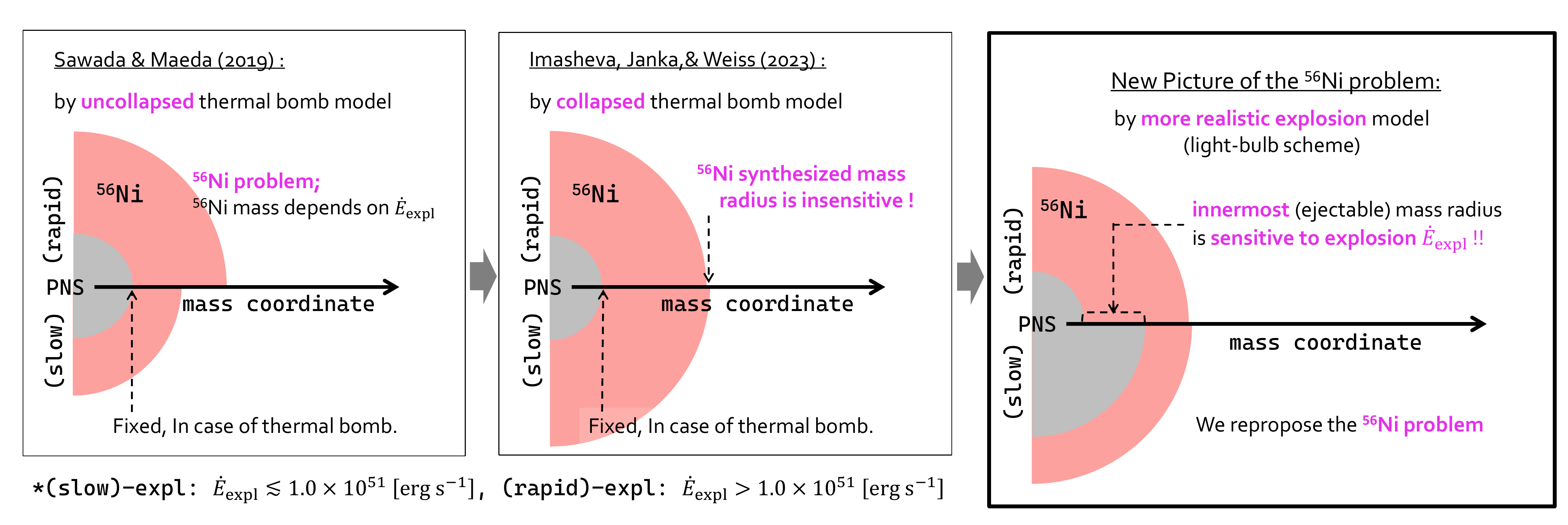}
  \caption{
  Schematic picture to the `\nickel problem in CCSNe' as suggested by the the previous studies and this study, respectively.
  At first, \sawadat~raised the $^{56}$Ni problem because the $^{56}$Ni synthesized region varies with the growth rate of the explosion energy \edot~when an un-collapsed progenitor is used. 
  When taking into account that the progenitor collapses just before the explosion, the $^{56}$Ni synthesized region becomes insensitive to \edot, and thus, \prevt~proposed a disappearance of the $^{56}$Ni problem. 
  However, this study re-proposes the $^{56}$Ni problem on the grounds that while the \nickel~synthesized region is insensitive to \edot, the ejectable innermost mass radius depends on the \edot, as calculated using the light-bulb scheme in which the PNS masses is determined self-consistently.}
  \label{fig:image}
 \end{center}
\end{figure*}

   \item 
{\bf[Comparison to \prevt~]}\\
Finally, we compare our results with those of \prevt~who just recently pointed out the most obvious doubts about the `\nickel problems'.
Figure \ref{fig:image} is a schematic comparison of our results with theirs.
Their argument is that the correlation between \edot~ and \nickel disappears when the initial collapse is included, and that further initial collapse inversely results in more \nickel being synthesized in slower explosions. 
Noting that the innermost ejecta radius is fixed in the thermal bomb model, their argument is consistent with the present results where \nickel is synthesized to almost the same mass coordinate independent of \edot, shown in Figure \ref{fig:comp-shock}.
We also confirm that \nickel is synthesized slightly more outwardly in models with slower initial collapse (i.e., the $M_\mathrm{ZAMS}=19.5 M_\odot$ model).
The difference from their study is the treatment of the innermost mass coordinate of the ejecta, i.e., their inner boundary condition.
Our explosion model determines the innermost mass coordinate of the ejecta self-consistently. 
We then found that the innermost material that could be ejected in the high-\edot~model could not achieve the escape condition in the low-\edot~model, leading to moving the innermost mass coordinate of the ejecta outward.
In fact, for low \edot~models, \prevt~ themselves had mentioned the possibility that some of the innermost material may be unable to achieve escape conditions, remain gravitationally bound, and thus not contribute to the yield, and we confirmed this in this paper.
Although our results are consistent with theirs, we confirm that \nickel~problems reappear because the innermost ejecta radius shifts depending on the intensity of the \edot.
\end{enumerate}

We conclude that the modern slow explosion ($\dot{E}_\mathrm{expl}\lesssim1$ Bethe s$^{-1}$ ) can reproduce the observations of a standard Type II supernova. 
However, this is only a statement of a principal possibility. 
How much \nickel can be synthesized is an important benchmark for multidimensional self-consistent simulations, and it should be confirmed whether the median value for a standard Type II supernova ($\approx0.03M_\odot$) is indeed achieved.

On the other hand, the \nickel problem clearly exists in the explosion mechanism of SE-SNe, that is, the modern slow explosions cannot reproduce the SE-SNe observations.
As a simple and straightforward solution that satisfies the \nickel problem without fine-tuning, we conclude that the SE-SNe favors active explosions in the early stages of shock revival ($\dot{E}_\mathrm{expl}\gtrsim2$ Bethe s$^{-1}$).
Since such high explosion energies are probably inconsistent with the standard explosion mechanism, the \nickel problem may require a different explosion mechanism for the SE-SNe.
\citet{2019A&A...628A...7A} had already suggested from observations, but our results once again imply significant differences in the progenitor structures and/or the explosion mechanism between type II and SE-SNe.

\section{Summary}\label{sec:summary}
In this paper, we investigated the effect of the explosion energy growth rate \edot~ on the behavior of \nickel nucleosynthesis in CCSNe.
For numerical simulations, we employed the 1D Lagrangian hydrodynamic code in which neutrino heating and cooling terms are taken into account by the light-bulb approximation. 
The initial conditions are taken from \citet{2018ApJ...860...93S}, which have $M_\mathrm{ZAMS}= 12.3, 16.0, 18.0,$ and $19.5 M_\odot$. 

Our first purpose was to present a detailed picture of \nickel nucleosynthesis in CCSNe with self-consistent explosion modeling.
We found that \nickel is synthesized up to the almost same mass coordinate independent of \edot.
We also found that in the low-\edot~model, some of the innermost material that was ejected in the high-\edot~model failed to achieve the escape condition, leading to moving the innermost mass coordinate of the ejecta to the outer positions.
This means that while the \nickel nucleosynthesis volume is insensitive to the nature of the explosion, the ejected amount of \nickel is highly dependent on how much of the innermost PNS surface region is ejectable.

Furthermore, our other goal was to sort out the recent controversial \nickel problem.
We found that there is a decreasing trend of the synthesized amount of \nickel toward decreasing \edot.
Compared to observations, we found that the modern slow explosion ($\dot{E}_\mathrm{expl}\lesssim1$ Bethe s$^{-1}$ ) can reproduce the observations of a standard Type II supernova in a principal.
However, this does not mean that the \nickel problem has been solved, and the \nickel synthesis ($M_\mathrm{Ni}\approx0.03M_\odot$) still remains an important benchmark for multi-D self-consistent simulations.
It should be checked whether they are truly achieved.
And extremely important are the comparison results with SE-SNe.
We found that the median value of SE-SNe ($M_\mathrm{Ni}\approx0.07M_\odot$) is challenging to reproduce in the modern slow explosion ($\dot{E}_\mathrm{expl}\lesssim1$ Bethe s$^{-1}$ ).
As a simple and straightforward solution that satisfies the amount of \nickel without fine-tuning, the SE-SNe favors active explosions in the early stages of shock revival ($\dot{E}_\mathrm{expl}\gtrsim2$ Bethe s$^{-1}$).
Thus, we conclude that there are significant differences in the progenitor structures and/or the explosion mechanism between type II and SE-SNe.

\software{{\tt blcode}, {\tt torch}\citep{1999ApJS..124..241T}}

\section*{Acknowledgments}
The work has been supported by Japan Society for the Promotion of Science (JSPS) KAKENHI grants 21J00825, 21K13964 (RS), 18H05437,  20H00174, 20H01904, and 22H04571 (YS).

\begin{center}
\begin{table*}
\centering
\caption{Summary of simulations} \label{tbl:result}
  \begin{tabular}{rcc|cccccc} \hline\hline
    $M_\mathrm{ZAMS}$
    &$M_{s/k_b=4}$ 
    \tablenote{Mass Coordinate with $s = 4k_B$ baryon$^{-1}$.} 
    &$\xi_{s/k_b=4}$ 
    \tablenote{Compactness parameter of $M_{s/k_b=4}$ which is defined as $(M_{s/k_b=4}/M_\odot)/(R_{s/k_b=4}/1000\mathrm{km})$.} 
    &$L_{\nu_e}$ &$\dot{E}_\mathrm{expl}$ 
    \tablenote{Neutrino luminosity.}
    &$M_\mathrm{PNS}$ 
    \tablenote{PNS mass which we define as the mass coordinate that experienced $r<200$ km as the neutrino-driven wind and separate it from the ejecta.} 
    &$M_{T_9=5}$ 
    \tablenote{Mass Coordinate of shock at a time when the post-shock temperature is $T = 5 \times 10^9$ K.} 
    &$M_r(X_\mathrm{^{56}Ni}>0.5)$  
    \tablenote{Outermost mass radius where the \nickel is $X_\mathrm{^{56}Ni}>0.5$ synthesized.} 
    &$M_\mathrm{^{56}Ni}$ 
    \tablenote{\nickel mass computed in a post-process with the 640-isotope.} \\
    
    $[M_\odot]$&$[M_\odot]$&&$[10^{52}$erg s$^{-1}]$&$[10^{51}$erg s$^{-1}$] &$[M_\odot]$ &$[M_\odot]$ &$[M_\odot]$ &$[M_\odot]$ \\ \hline \hline
    
    $12.3$&$1.462$ & $0.837$
    & 1.4& $0.46$&  $1.464$ & $1.514$& $1.523$&  $0.049$\\ 
    &&& 2& $0.60$&  $1.464$ & $1.509$& $1.516$&  $0.047$\\ 
    &&& 3& $1.04$&  $1.458$ & $1.507$& $1.512$&  $0.049$\\  
    &&& 4& $1.75$&  $1.449$ & $1.507$& $1.512$&  $0.057$\\  
    &&& 5& $2.88$&  $1.427$ & $1.510$& $1.514$&  $0.078$\\  \hline
    
    $16.0$&$1.618$ & $0.700$
     & 2& $0.48$&  $1.617$ & $1.655$& $1.659$& $0.039$\\ 
    &&& 3& $0.82$&  $1.615$ & $1.651$& $1.655$& $0.039$\\ 
    &&& 4& $1.35$&  $1.605$ & $1.650$& $1.653$& $0.045$\\ 
    &&& 5& $2.39$&  $1.582$ & $1.653$& $1.671$& $0.068$\\ 
    &&& 6& $3.47$&  $1.561$ & $1.651$& $1.656$& $0.087$\\ 
    &&& 7& $4.62$&  $1.536$ & $1.650$& $1.655$& $0.109$\\  \hline

    $18.0$&$1.535$& $0.793$
      & 2& $0.83$&  $1.537$ & $1.593$& $1.607$&  $0.062$\\ 
    &&& 3& $1.06$&  $1.537$ & $1.593$& $1.600$&  $0.059$\\ 
    &&& 4& $1.49$&  $1.535$ & $1.593$& $1.599$&  $0.060$\\  
    &&& 5& $2.12$&  $1.530$ & $1.593$& $1.600$&  $0.065$\\  
    &&& 6& $3.10$&  $1.514$ & $1.595$& $1.602$&  $0.081$\\ 
    &&& 7& $4.23$&  $1.514$ & $1.608$& $1.615$&  $0.112$\\ \hline

    $19.5$&$1.714$& $0.762$
     & 2& $0.70$&  $1.912$ & $1.962$& $1.970$&  $0.054$\\  
    &&& 3& $1.22$&  $1.720$ & $1.806$& $1.816$&  $0.090$\\ 
    &&& 4& $1.64$&  $1.717$ & $1.801$& $1.811$&  $0.089$\\  
    &&& 5& $2.10$&  $1.714$ & $1.800$& $1.810$&  $0.092$\\  
    &&& 6& $2.80$&  $1.709$ & $1.801$& $1.810$&  $0.097$\\  
    &&& 7& $3.69$&  $1.691$ & $1.801$& $1.810$&  $0.114$\\  \hline
\hline
  \end{tabular}
\end{table*}
\end{center}

\bibliography{ref}{}
\bibliographystyle{aasjournal}

\appendix
\section{\nickel Synthesis Condition}\label{sec:nickel}

Here, we briefly review the general features of \nickel nucleosynthesis in CCSNe, which is classically well enough understood \citep[e.g.,][]{1968ApJS...16..299B,1973ApJS...26..231W,1999ApJ...511..862H}.
To illustrate the condition of \nickel synthesis, for simplicity, we use parameterized adiabatic expansion profiles.
Let us focus on a certain Lagrangian particle. 
Assuming that a passing shock wave heats and compresses the particle to peak temperature $T_0$ and peak density $\rho_0$, and then it expands and cools in a constant $T^3/\rho$ track, we can write the temperature and density evolution \citep{1973ApJS...26..231W} as
\begin{align}
T(t)&= T_0 \exp{(-t/3\tau_\mathrm{dyn.})},~~\nonumber\\
\rho(t) &= \rho_0 \exp{(-t/\tau_\mathrm{dyn.})}~, \label{eq:ad-ex}
\end{align}
where the expansion timescale of the ejecta $\tau$ is the following \citep{1964ApJS....9..201F};
\begin{equation}
    \tau_\mathrm{dyn.}= (24\pi G \rho_0)^{-1/2}\approx 446/\rho_0^{1/2}~ \mathrm{sec}.
\end{equation}
This expansion trajectory reproduces the general temperature and density trajectories in spherically symmetric or multi-dimensional CCSNe explosions, especially in the explosive nucleosynthesis region  \citep[e.g.,][]{2010ApJS..191...66M,2011ApJ...741...78M,2015ApJ...807..110J}.
We calculate a reaction network of 640 nuclear species with {\tt torch} code \citep{1999ApJS..124..241T}, with the initial composition $X(^{28}\mathrm{Si})=1$.
We then check the behavior of \nickel nucleosynthesis using expansion profiles with peak temperature $T_0$ and peak density $\rho_0$ as parameters.

\begin{figure}
\centering
  \includegraphics[width=0.49\textwidth]{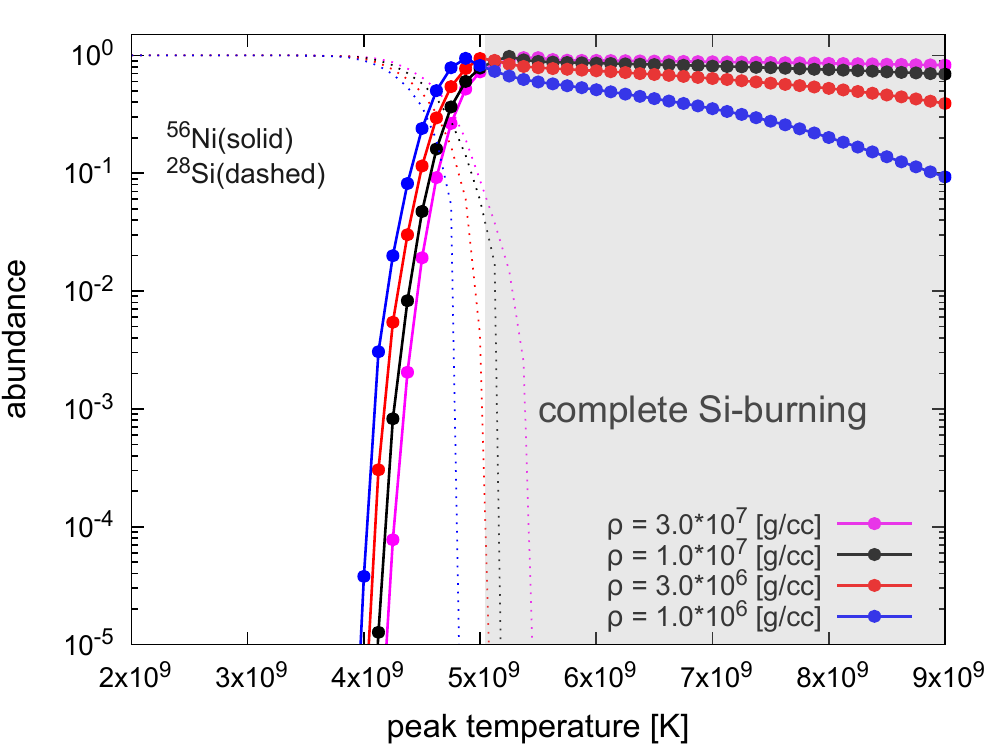}
\caption{The abundance of synthesized \nickel as a function of peak temperature $T_0$ with the peak density $\rho_0$ at $Y_e=0.5$, calculated using the adiabatic expansion profile of Eq. \eqref{eq:ad-ex}.} 
\label{fig:ni-prof}
\end{figure}

Figure \ref{fig:ni-prof} shows the abundance of synthesized \nickel as a function of the peak temperature $T_0$ in the adiabatic expansion profile.
When the peak temperature is $T_9\gtrsim5$ ($T_9\equiv T/10^9$ K), we can see that a sizable amount of \nickel is synthesized, almost independent of the peak density $\rho_0$.
In this temperature region, $^{28}$Si is wholly depleted at the same time as satisfying $^{56}$Ni synthesis, so this explosive nucleosynthesis is referred to as `complete Si burning'.
This depletion of $^{28}$Si occurs when the lifetime over which $^{28}$Si burns up is much shorter than the given expansion time scale, i.e., $\tau(\mathrm{^{28}Si})\ll\tau_\mathrm{dyn.}$.
According to \citet{1973ApJS...26..231W}, to estimate this condition quantitatively, we define the depletion point of $^{28}$Si as when its mass fraction falls below about 0.005 (i.e., $X_f(\mathrm{^{28}Si}) < 0.005$). Then, complete Si combustion is estimated to occur under the following initial conditions;
\begin{equation}
    T_9 \gtrsim 5.0\times\left(\cfrac{\rho_0}{10^6~\mathrm{g~cm^{-3}}}\right)^{1/68}
\end{equation}
This implies that an initial temperature must exceed $T_9 \approx 5$ to promote the depletion of $^{28}$Si in order for the complete silicon burning to occur at a reasonable density.
This estimate is roughly consistent with our numerical result in Figure \ref{fig:ni-prof}.

In complete Si burning, the forward and reverse reactions in the main reaction channel proceed much faster than the time scale of the change in thermal conditions. 
The compositions thus basically follow the Nuclear Statistical Equilibrium (NSE), except for the slow triple-alpha process.
Then, as the temperature decreases rapidly, the reaction rates decrease and the abundance pattern `freezes-out' \citep{1999ApJ...511..862H}.
In NSE, the composition ratio is determined to minimize the Helmholtz free energy ($\mathcal{F} = (U - \mathcal{Q}) - TS$) for the nuclide mass fraction \citep{1965MmRAS..69...21C}.
When the temperature is not too high ($T_9<10$), the binding energy per nucleon (BEN) $\mathcal{Q}=B/A$ roughly determines the abundance pattern of NSE.
In environments where the number of protons and neutrons is almost equal ($Y_e\approx 0.5$), the most abundant isotope in NSE is $^{56}$Ni.
Therefore, the main synthesis process of \nickel in the SNe explosion is not a two-/three-body reaction, but a `freezes-out' of the NSE abundance pattern that occurred via the `photodisintegration-rearrangement reactions' associated with the photodisintegration of $^{28}$Si.

\begin{figure}
\centering
  \includegraphics[width=0.49\textwidth]{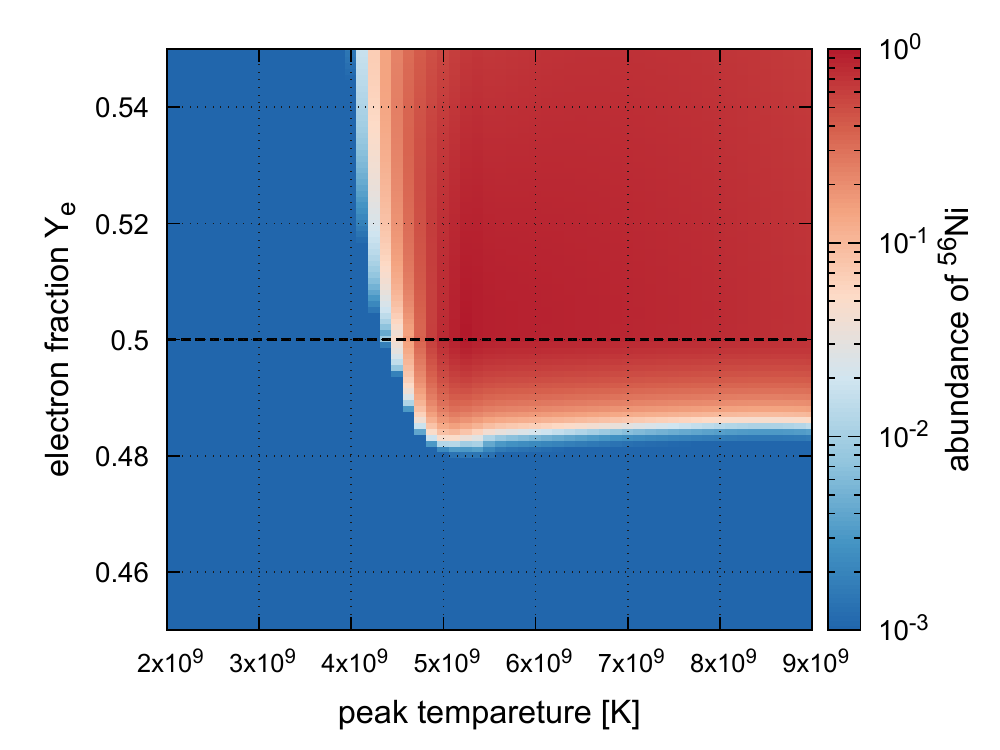}
\caption{The abundance of synthesized \nickel in the peak temperature $T_0$ and the initial electron fraction $Y_e$ plane at the peak density $\rho_0=10^7$ g cm$^{-3}$, calculated using the adiabatic expansion profile of Eq. \eqref{eq:ad-ex}.} 
\label{fig:ni-prof-ye}
\end{figure}

Next, Figure \ref{fig:ni-prof-ye} shows 
what the \nickel synthesis would be in a different environment for $Y_e$ with different peak temperatures.
We can clearly see that while \nickel synthesis in the proton-rich environment ($Y_e> 0.5$) is relatively the same as that in the $Y_e\approx 0.5$ environment, its synthesis is quite suppressed in the neutron-rich environment ($Y_e< 0.5$).
As explained by \citet{2008ApJ...685L.129S}, $Y_e$ dependence of nickel nucleosynthesis can be understood qualitatively by focusing on the BEN, $\mathcal{Q}=B/A$, and the Helmholtz free energy $\mathcal{F}$.

\begin{figure}
\centering
  \includegraphics[width=0.49\textwidth]{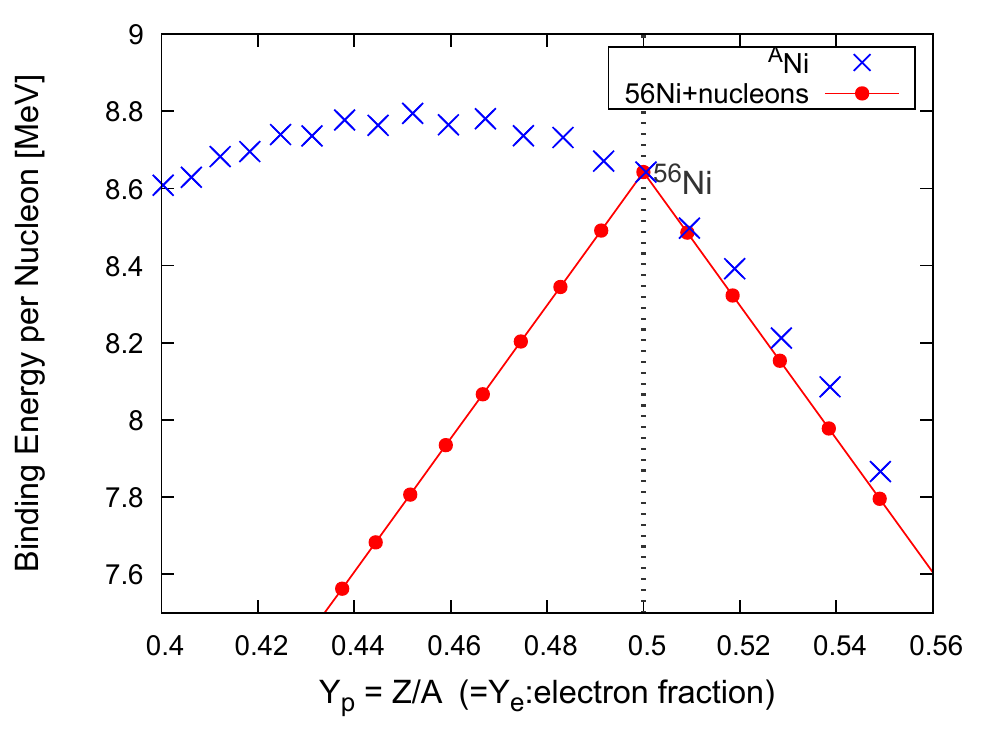}
  \includegraphics[width=0.49\textwidth]{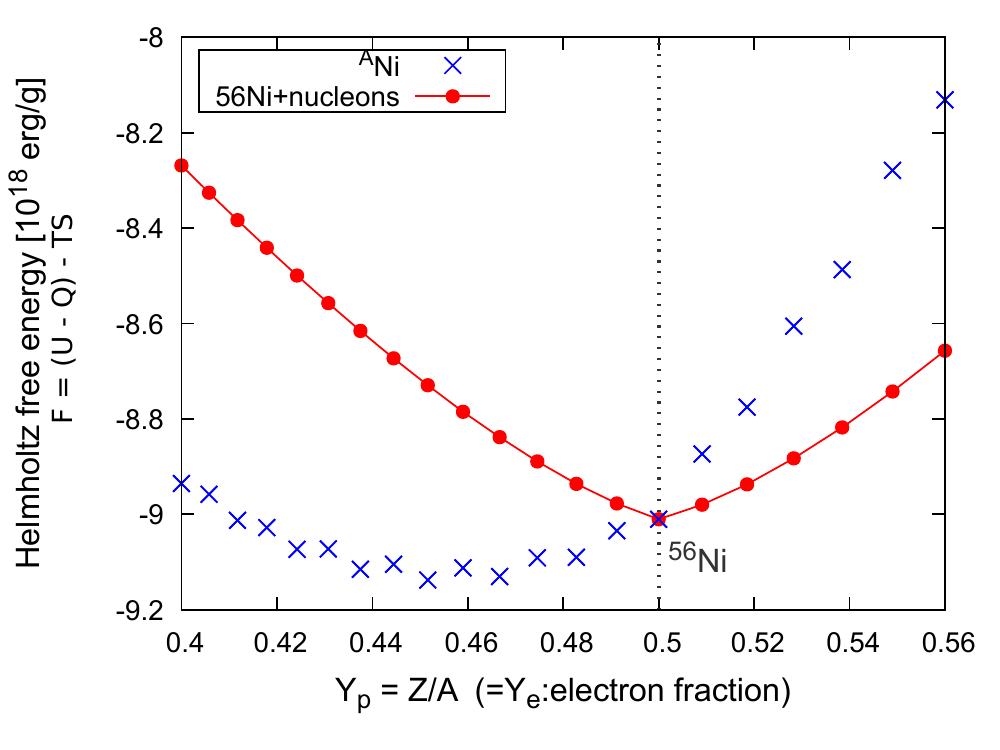}
\caption{Binding energy per nucleon (left) and the Helmholtz free energy per nucleon (right) with the number of protons per nucleon $Y_p = Z/A$ (corresponds to $Y_e$, and $28/Y_e$ corresponds to the mass number $A$ of each Ni isotope) on the horizontal axis.
The blue crosses show the binding energy per nucleon (BEN) of the Ni-isotope, and the red circles indicate the mean BEN of a mixed composition, which is mixed with \nickel and an appropriate number of free protons or free neutrons for any given $Y_e$ (\nickel+nucleons).
The mean BEN of the mixed composition is derived by $\bar{\mathcal{Q}}=\Sigma_i~\mathcal{Q}_in_i/n_B$.
The Helmholtz free energies were calculated using Helmholtz EoS \citep{2000ApJS..126..501T} with $T=5\times10^9$ K and $\rho=10^7$ g cm$^{-3}$ fixed.} 
\label{fig:ni-ben}
\end{figure}

Figure \ref{fig:ni-ben} (left) illustrates the BEN of various Ni-isotopes with blue crosses, where the number of protons per nucleon  ($Y_p = Z/A$, which corresponds to $Y_e$) on the horizontal axis.
To analyze stability, we consider a mixed composition of \nickel~plus an appropriate number of free protons or free neutrons (\nickel+nucleons) to match any $Y_e$.
The mean BEN of the mixed composition is calculated as $\bar{\mathcal{Q}}=\Sigma_i~\mathcal{Q}_in_i/n_B$, where $\mathcal{Q}_i$ is the binding energy of the nucleus $i$ and $\mathcal{Q}_\mathrm{neut}=\mathcal{Q}_\mathrm{prot}=0$. 
By comparing the BENs for each $Y_e$, we find that the single Ni-isotope is significantly more stable than the \nickel +nucleons mixture in the neutron-rich region ($Y_e < 0.5$), but not in the proton-rich region ($Y_e \geq 0.5$). 
While this picture is somewhat modified when entropy is taken into account, this simple discussion shows that \nickel~synthesis is suppressed in the neutron-rich region.

Figure \ref{fig:ni-ben} (right) shows Helmholtz free energies ($\mathcal{F} = (U - \mathcal{Q}) - TS$) for a single composition of only Ni isotopes and, as before, for a mixed composition of \nickel plus an appropriate number of free protons or free neutrons.
Helmholtz free energy was calculated using Helmholtz EoS \citep{2000ApJS..126..501T} with $T=5\times10^9$ K and $\rho=10^7$ g cm$^{-3}$ fixed.
In an environment with the same $T$ and $\rho$, the internal energy can be considered to be almost constant, $U\approx\mathrm{Const}$. 
The BEN term, $\mathcal{Q}$, then contributes to the Helmholtz free energy on the order of $1$ MeV/nuclear $\approx \mathcal{O}(10^{18})$ erg g$^{-1}$. 
For the $TS$ term, the entropy is increased by the free nucleons in the \nickel + nucleon mixture composition compared to the single Ni-isotope composition. It thus works toward lowering the Helmholtz free energy on the order of $\mathcal{O}(10^{17})$ erg g$^{-1}$.
Therefore, in the proton-rich region ($Y_e \geq 0.5$), the difference in BEN between the \nickel +nucleon mixture and the single Ni-isotope is $\Delta\mathcal{Q}\lesssim \mathcal{O}(10^{17})$ erg g$^{-1}$, which is not much larger than the correction by entropy enhancement. 
As a result, the correction for the entropy enhancement due to the free nucleon makes the \nickel +nucleon mixture more stable.
On the other hand, in the neutron-rich region ($Y_e < 0.5$), the difference in BEN is $\Delta\mathcal{Q}\sim \mathcal{O}(10^{18})$ erg g$^{-1}$, which is sufficiently larger than the effect of the correction by entropy enhancement, and thus the single Ni-isotope composition is more stable, as predicted from the BEN behavior.

To summarize the above discussion, in the temperatures region where $T_9\gtrsim5$ are attained behind the shock, NSE is achieved except for the slow triple-alpha process. 
Thus, the abundance pattern depends only on $Y_e$ and the entropy, and is dominated by iron-group elements \citep[e.g.,][]{1999ApJ...511..862H}. The most abundant element in this region of complete Si burning with alpha-rich freeze-out is \nickel, provided $Y_e\gtrsim0.5$.
The \nickel synthesis conditions in CCSNe explosion are an environment where the peak temperature $T_9\gtrsim5$ is experienced by the shock and the electron fraction of the material is $Y_e\gtrsim0.5$.

\end{document}